\pdfoutput=1
\documentclass[final,1p,times]{elsarticle} 
\usepackage{graphicx} 
\usepackage{amssymb} 
\usepackage{amsthm} 
\usepackage{lineno} 

\newcommand{\mean}[1]{\left\langle #1 \right\rangle} 
\newcommand{\ptopi}{$p/\rm \pi$ }

\newcommand{\ptopip}{$p/\rm \pi^{+}$ }

\newcommand{\rootsnn}[1]{$\sqrt{s_{_\mathrm {NN}}} = #1$~GeV}
\newcommand{\roots}[1]{$\sqrt{s} = #1$~GeV}
\newcommand{\pt}{$p_{T}$}

\newcommand{\pT}{$p_{T}$}
\newcommand{\meanpt}{$\left\langle  p_T \right\rangle$}

\newcommand{\piplus}{$\pi^{+}$}
\newcommand{\piminus}{$\pi^{-}$}

\newcommand{\flowvtwo}{${\rm v}_2$}

\newcommand{\sNN}{$\sqrt{s_{_\mathrm{NN}}}$ }

\journal{Nuclear Physics A} 
\begin{document} 

\begin{frontmatter} 


\title{Overview and Recent Results from BRAHMS}

\author{F.Videb{\ae}k  for the BRAHMS collaboration}

\address{Physics Department, Brookhaven National Laboratory}

\begin{abstract} 
The BRAHMS experiment was designed to measure and characterize in particular the properties
of rapidity dependence of particle production in heavy ion collisions. 
The data-taking is now over,
results of several years of analysis have been published and demonstrates several important features
of the rapidity dependence, not envisioned from the start of the RHIC program.
The bulk properties of the system formed at high rapidity resemble that of systems at lower energies at mid-rapidity when referenced via the baryo-chemical potential. 
New physics in AA are essentially observed at mid-rapidity including the demonstration that high-\pT~ suppression is a final state effect.
Another key result  is that in d+A collisions at forward rapidities
where the very low-x region of the nucleus was probed, a strong suppression of pion
production was observed consistent with the picture of gluon saturation.
The latest results examines the centrality and rapidity dependence of nuclear stopping, the particle production of pions, collective expansion vs. rapidity, and the baryon enhancement at intermediate values of \pT .
\end{abstract} 

\end{frontmatter} 


\section{Introduction}

In these proceedings I review a  number of BRAHMS results in the perspective of the early goals and expectations for the experiment, and what has been achieved over the 6 RHIC runs in which we took data.
In addition I discuss a number of unforeseen results that came from later development of the RHIC physics and the open-minded approach to experimental research. 
This contribution does neither give a the full history, nor all results, for which I refer to our extensive publications  in journals and conference proceedings. See e.g. Ref.~\cite{BRAHMS:WhitePaper} for early results and discussion.

The experiment was proposed in 1990 and the Technical Design Report was approved in 1997. 
A partial spectrometer system was ready for the first RHIC run, and both spectrometers were completed for the second RHIC run. BRAHMS took data for 6 RHIC runs with the last data set  recorded during the 62.4 GeV p+p run in 2006.

The initial goals of BRAHMS were to measure identified particle production over a wide range of rapidity, \pT, and for a range of collisions systems A+A, p+A and p+p. This was to clarify the reaction mechanism and look for possible effects of the QGP that might be visible in inclusive spectra.
Additional goals were developed during the construction process and the early RHIC running leading to the important d+Au run in 2003, and a successful BRAHMS transverse spin program (see e.g.~\cite{BRAHMS:ANN_62gev}).
BRAHMS took heavy ion data at \sNN= 130, 200 and 62.4 GeV in Au+Au, and Cu+Cu collisions.
Important reference data was recorded in d+Au and p+p at 200 GeV, and  p+p spin data taken at 200 and 62.4 GeV.
The setup of the BRAHMS experiment is described in detail in Ref.~\cite{BRAHMS:NIM}.

\section{Net-baryons in AA and pp}

It  is important to understand nuclear stopping in detail since it is a requisite for
the formation of dense systems in describing the conversion of the initial kinetic 
energy into matter excitation at mid-rapidity. 
A detailed description of how this transport
takes place is a necessary ingredient in our overall understanding of the reactions and in the
expectations of forming and for studying the properties of QGP.
\begin{figure}[ht]
\begin{center}
\includegraphics[width=2.5in,angle=90]{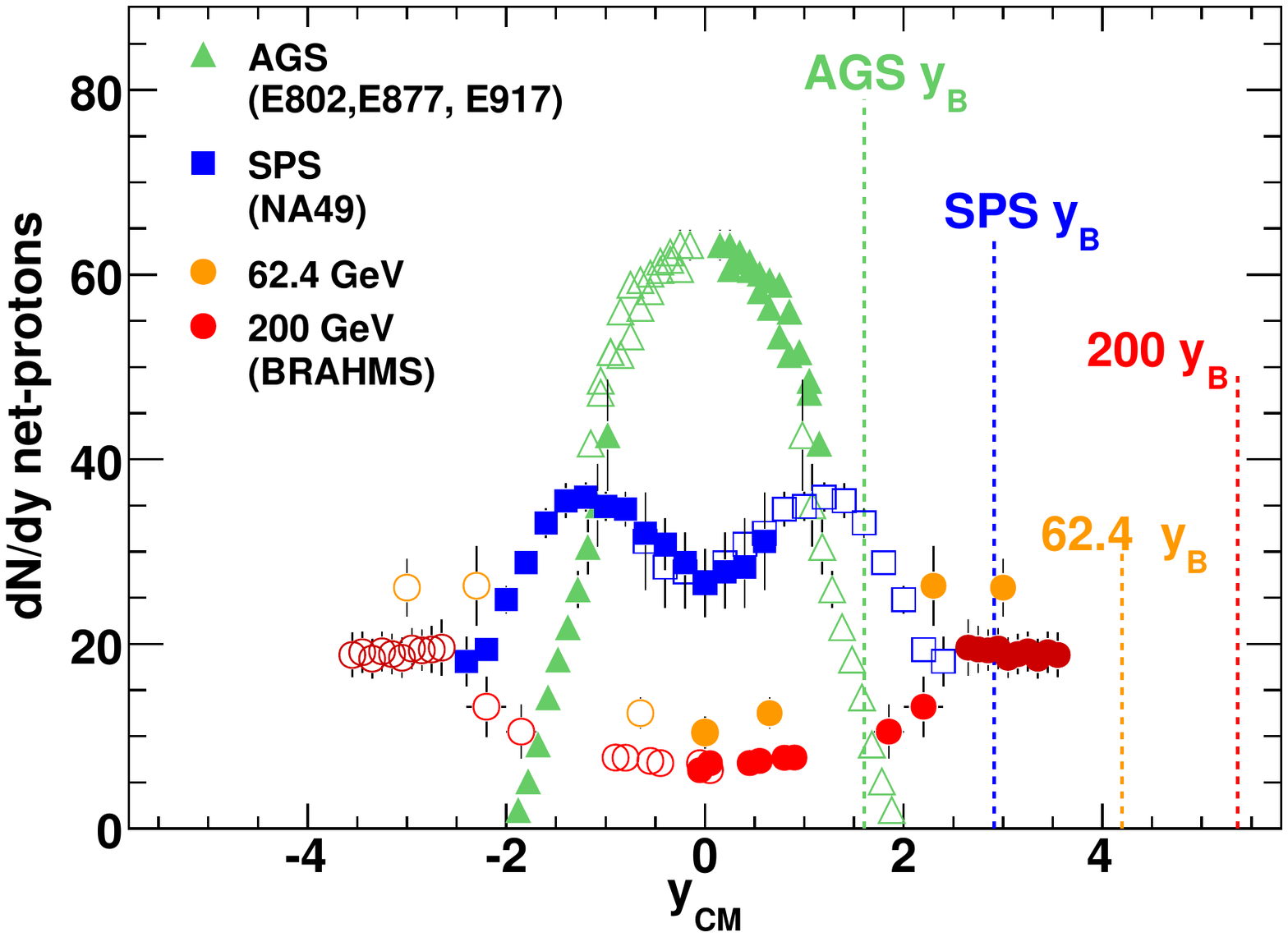}
\end{center}
\vspace{-0.8cm}
\caption{(color online) Net-proton distributions from AGS to top RHIC energies. The high-rapidity 200 GeV data points are preliminary.
The beam rapidity $y_B$  at each energy is indicated by the dashed lines.}
\label{fig:netp}
\end{figure}
The second question is whether the picture by Bjorken ~\cite{Bjo83} that proposes a transparent scenario where  the central region is net-baryon free, and most baryons are at high rapidity 
after the collision is achieved at the highest energies. 
At the lower AGS energies rapidity distributions are closer to a Landau description where most 
of the baryons end up near mid-rapidity ~\cite{Vid95}.
This topic was addressed in BRAHMS via measurement of the net-proton distributions at 200 and 62.4 GeV.
In Fig.~\ref{fig:netp} we show the data from Ref.~\cite{BRAHMS:stop200} together with the new data at 62.4 GeV
~\cite{BRAHMS:stop62}. 
\begin{figure}[h]
\begin{center}
\includegraphics[width=2.5in,angle=90]{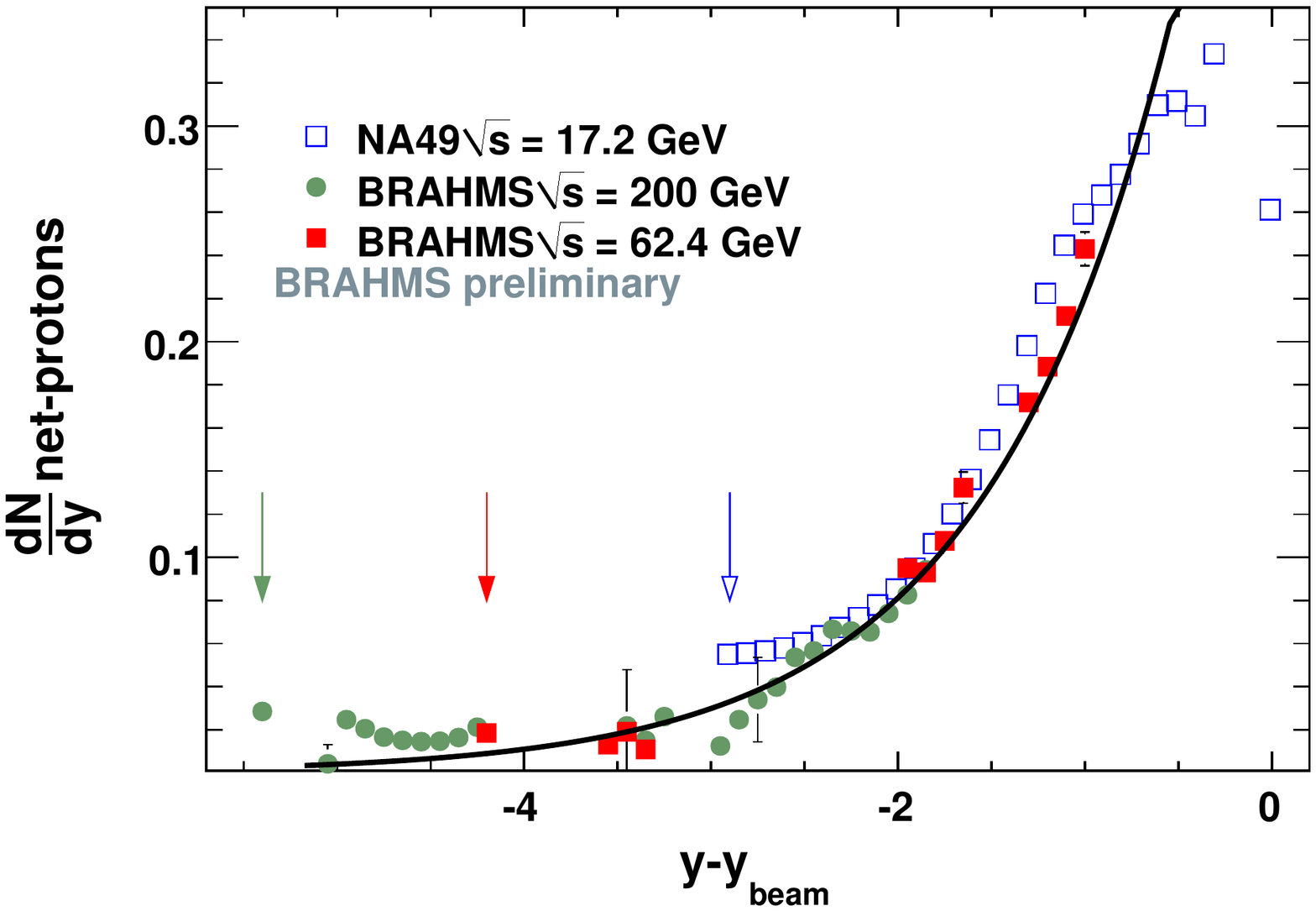}
\end{center}
\vspace{-0.8cm}
\caption{(color online)Net-proton rapidity distributions in p+p scaled to the beam rapidity. 
The arrows indicate the position of mid-rapidity for 200, 62.4 and 17.2 GeV, respectively. 
The NA49 data are from a recent preprint~\cite{Anticic:2009wd}.}
\label{fig:dndypp}
\end{figure}
The data from AGS to RHIC show a clear development of a net-proton poor region at mid-rapidity, with most baryons in the region less than 2 units away from beam rapidity. 
It is also demonstrated in another proceeding for this conference {~\cite{qm09HD} that the peripheral Au+Au collisions net-protons distributions are similar in shape to those of p+p.
At both energies we estimate that about 70\% of the incident energy is available for entropy production and longitudinal momentum of produced particles.

The average rapidity loss $\delta y$ is about 2 units at high energy for central collisions  in contrast to
p+p, where low-energy data give $\delta y \approx 0.6$, and with a near constant distribution in $dN/dx_F$ where $x_F$ is the Feynman x.
This implies that in rapidity space one will expect that $dN/dy = A e^{-(y-y_B)}$. In Fig.~\ref{fig:dndypp} we demonstrate that the data from NA49, and the BRAHMS RHIC p+p data from 62 and 200 GeV fall close to such an universal curve.
This common behavior  leaves little room for new mechanisms in p+p stopping up to RHIC energies.

\section{Meson Rapidity Distributions}
One of the key observations of BRAHMS is that the rapidity density distributions of produced particles, i.e., pions, kaons, and  anti-protons, exhibit a nearly Gaussian distribution as was shown in Ref.~\cite{BRAHMS:auaumeson}. 
Similar overall features have already been observed in central collisions of Au+Au at AGS and Pb+Pb at SPS.  
In Fig.~\ref{fig:meson} we show the preliminary data from run-4 for central, i.e., 0-10$\%$ together with the published data for $K^{+}$. The data have not been corrected for weak decay feed-down, a few percent correction. 
\begin{figure}[ht]
\begin{center}
\includegraphics[width=2.5in,angle=90]{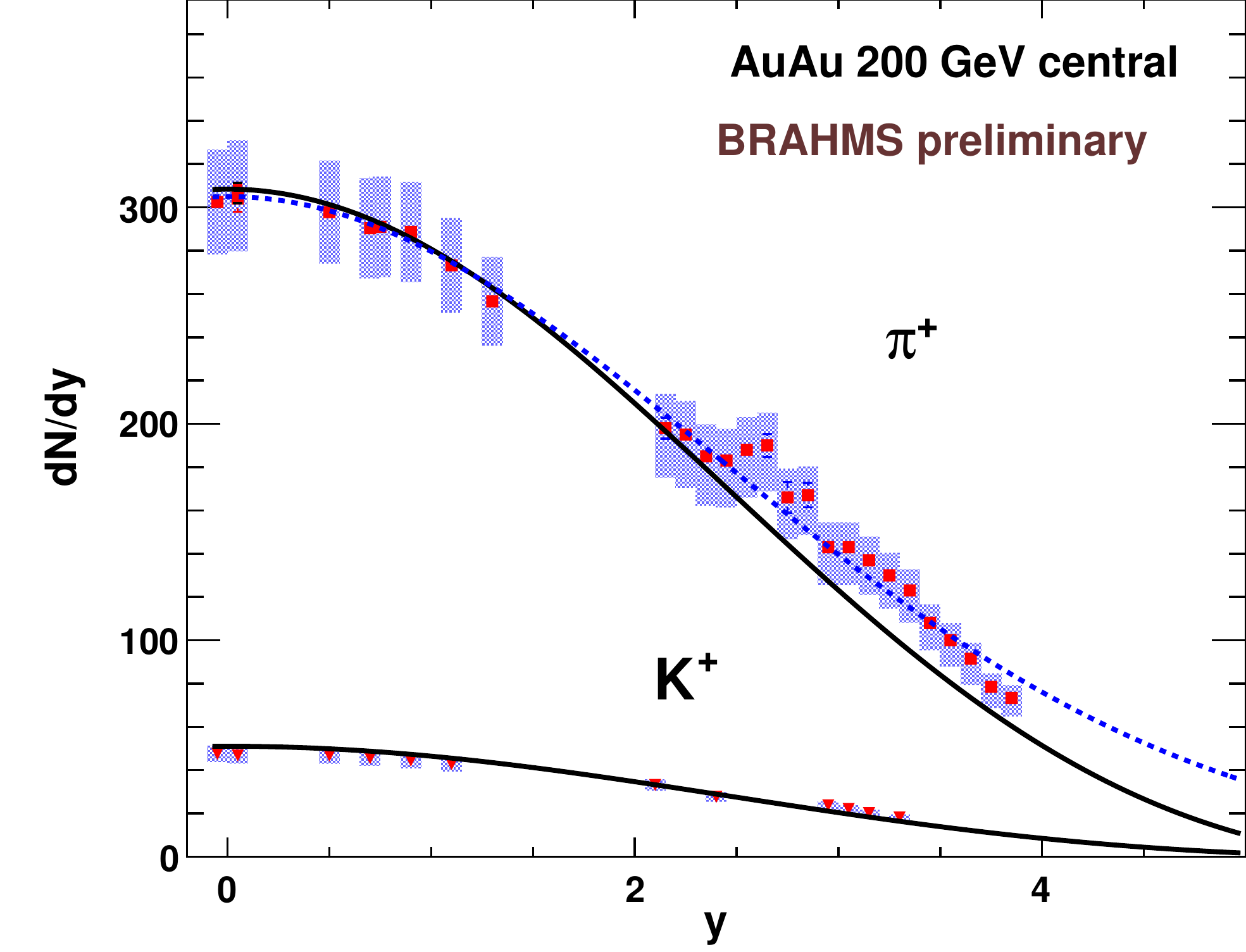}
\end{center}
\vspace{-0.8cm}
\caption{(color online) Rapidity distributions for pions and kaons in central Au+Au collisions at 200 GeV. 
The \piplus are for $0-10\%$ centrality, while the K$^{+}$ are the published data from Ref.~\cite{BRAHMS:auaumeson}}
\label{fig:meson}
\end{figure}
This is reminiscent of the hydrodynamical expansion model proposed by Landau, a feature that has attracted theoretical interest.
In a recent paper, Wong ~\cite{Wong:2008ex} derived the form $dN/dy \propto \exp{\sqrt{y_B^2-y^2}}$, instead of Landau's original distribution, dN/dy = $ \exp{\sqrt{L^2-y^2}}$, where $L\approx\ln{\sqrt{s_{\mathrm NN}}/2m_p}$.
In Fig.~\ref{fig:meson} we show the new form as a solid line  compared to the new preliminary data from the RHIC run-4.
The measurements give a slightly wider distribution than the modified Landau description, as indicated by the dashed curve which is a Gaussian description of the data. 
The interpretation of this is open; the agreement with the model does not necessarily indicate that the system developed full stopping.
\begin{figure}[ht]
\begin{center}
\includegraphics[width=2.5in,angle=90]{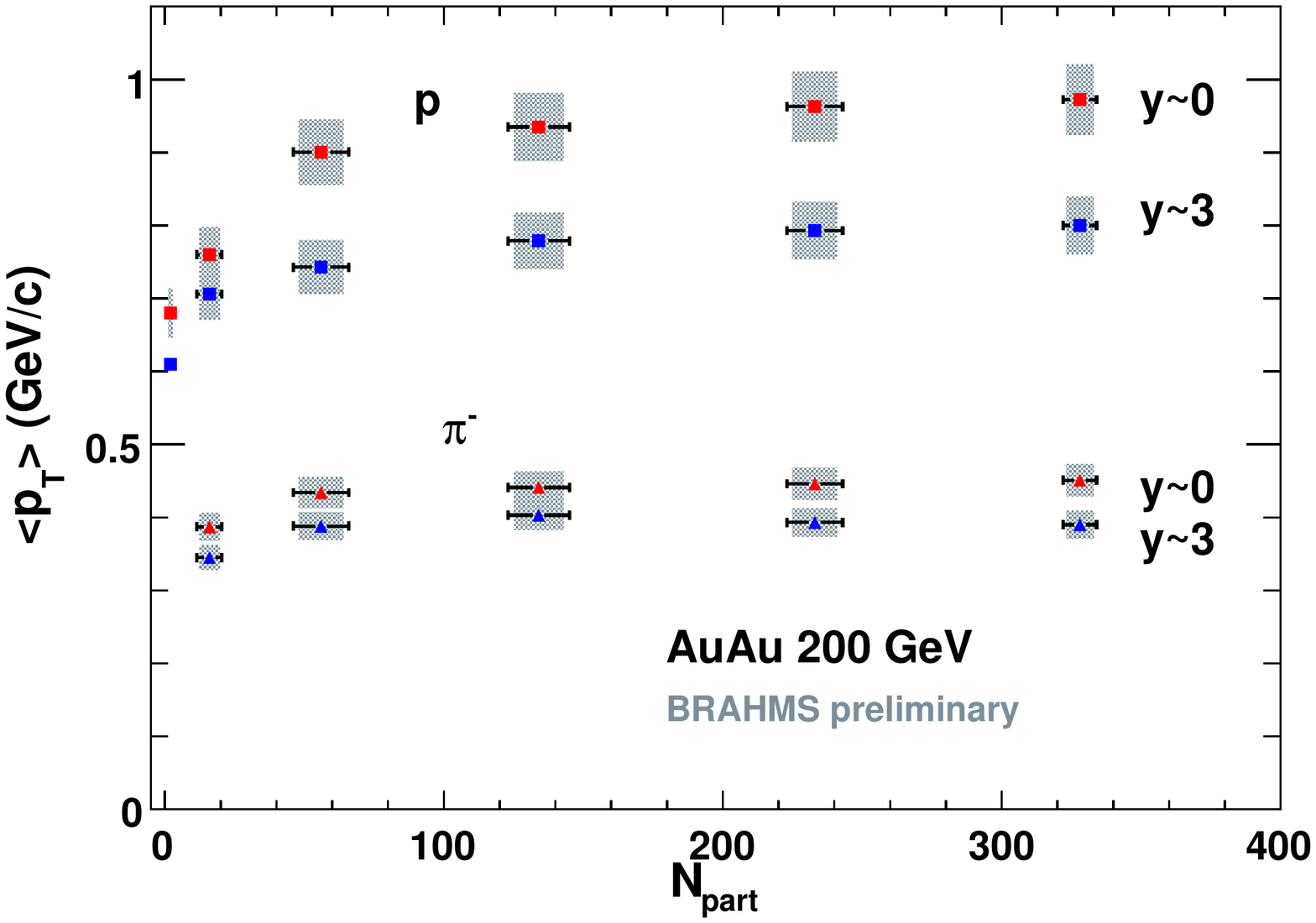}
\end{center}
\vspace{-0.8cm}
\caption{(color online) Mean \pT~vs. centrality for protons and pions and mid and forward rapidity in Au+Au
collisions at \rootsnn{200}.}
\label{fig:avpt}
\end{figure}

\section{Radial and Elliptic Flow}

The centrality dependence of the shapes of spectra, in particular heavier hadrons like protons, is an indication of the importance of radial flow in the collisions. In Fig.~\ref{fig:avpt} we show the dependence of the mean \pT~ for \piminus~ and protons at central and forward  (y = 3) rapidities.
The \meanpt~values are fairly constant in both rapidity and centrality for pions, whereas for protons they exhibit a rapid rise from the most peripheral collisions followed by a much slower rise. Their value at forward rapidities are clearly lower than at mid-rapidity. 
This change in \meanpt~is consistent with a reduction of radial flow at large rapidities.
Since the radial flow to a large degree develops in the later stages of the collision, it does not reflect the initial pressure.
On the other hand the elliptic flow is believed to be established in the early stages of the collision.
A strong azimuthal flow signature at RHIC suggests rapid system equilibration, leading to an almost perfect liquid state. 
The longitudinal extent of the flow behavior depends on the formation dynamics for this state and can be studied by measuring the pseudorapidity dependence of \flowvtwo, the second Fourier component of the azimuthal angular distribution. 
BRAHMS has measured for identified
particles $v_2$ as a function of \pT~(0.5  -2.0 GeV/$c$) for $0-25\%$ and  $25-50\%$ centrality
and pseudorapidity for \rootsnn{200}  Au+Au collisions.
These are discussed in more details in the contribution~\cite{qm09SJS} from this conference.
Figure \ref{fig:v2pt} shows $v_{2}(p_T/n_{q})$ plotted against the mean transverse energy per constituent quark,
 $\mean{E_T}/n_q$ for pseudorapidities of  $\eta \approx$  0, 1 and 3 for the $0-25\%$ centrality selection.
It is compared to the universal systematics at mid-rapidity  for scaled $v_2$ from R.~Lacey and Taranenko~\cite{Lacey}, and agrees well at $\eta=0$ and 1,
whereas $\eta=3$ is consistent with being lower than the systematics, a trend we clearly observe in the centrality range $25-50\%$ (not shown here, but in Ref.~\cite{qm09SJS} ).
This dependence with pseudo-rapidity together with the simultaneous reduction of  \pT, i.e., the spectral shape of the inclusive hadrons, makes these measurements of $v_{2}(p_T)$ consistent with the inclusive \flowvtwo~ vs. pseudo-rapidity  as observed by PHOBOS for charged hadrons~\cite{PHOBOS:v2}.

\begin{figure}[ht]
\begin{center}
\includegraphics[width=2.7in,angle=90]{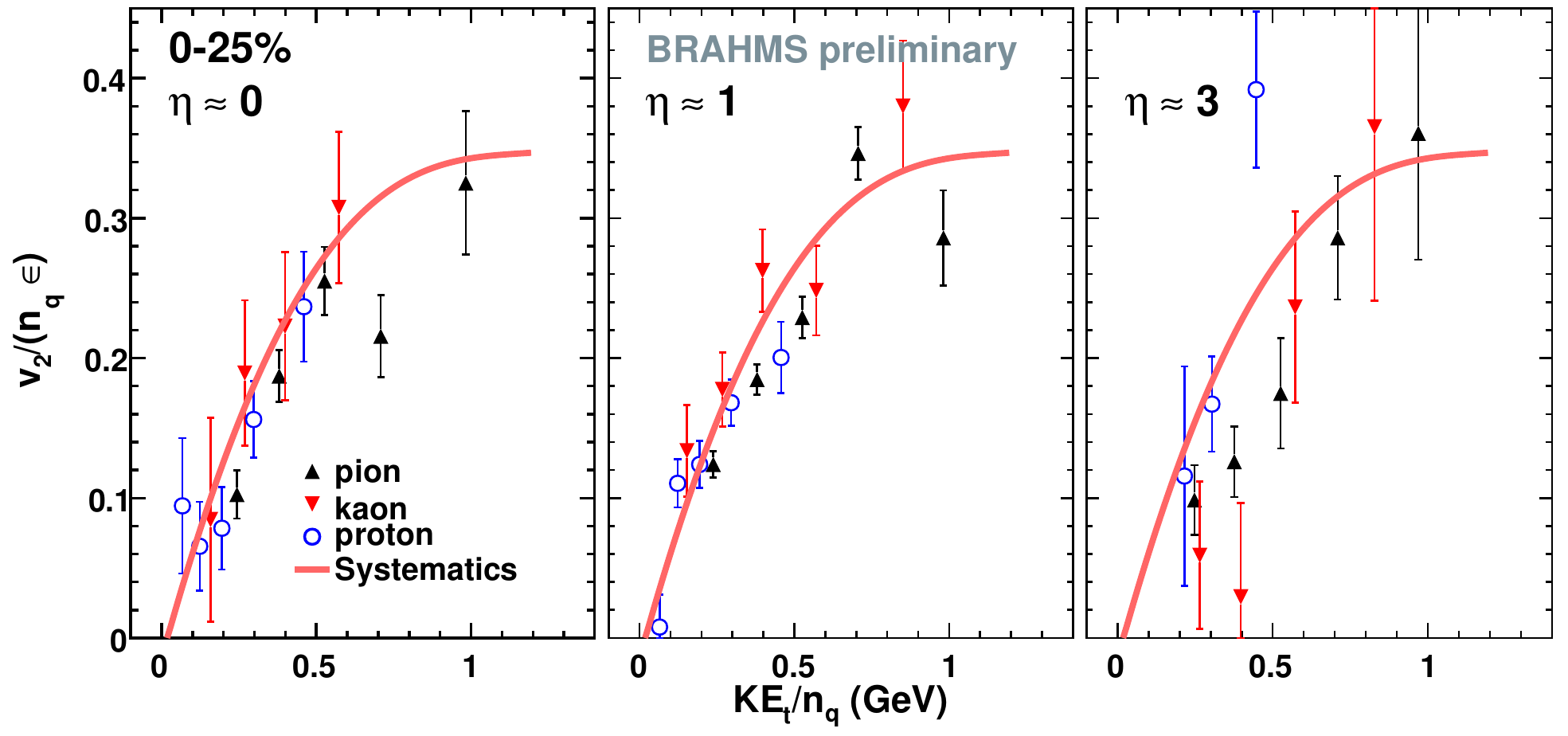}
\end{center}
\vspace{-0.6cm}
\caption{(color online) $v_2$(\pT) scaled by the number of valence quarks $n_q$  and the participant eccentricity as function of the transverse kinetic energy KE$_T$ scaled by  $n_q$ for $\eta \approx 0, 1 $ and 3.}
\label{fig:v2pt}
\end{figure}

\section{Baryon to Meson Ratio vs. Chemical Potential}
The discovery  of a large baryon to meson ratio at mid-rapidity in the \pT-range  $2 - 5$ GeV/$c$ at RHIC is taken as an indication that quark coalescence plays an important role for particle spectra.
The spectral shape and ratios reflect the underlying hadronization scenario (recombination vs. fragmentation), and the importance of radial flow of bulk medium. We would like to understand this in detail.
Energy and centrality  dependence of $p/\pi^{+}$ and ${\bar p}/\pi^{-}$  and their evolution on rapidity may allow us to verify the proposed scenarios. 
At large $\mu_{B}$, the
picture, suggested by mid-rapidity measurements, 
might be contaminated by final state hadron interactions, leading to 
a transition from the parton recombination scheme to
a hydrodynamical description that has a common velocity field for
baryons and mesons ~\cite{hirano, bronio}. 
\begin{figure}[ht]
\begin{center}
\includegraphics[width=2.7in,angle=90]{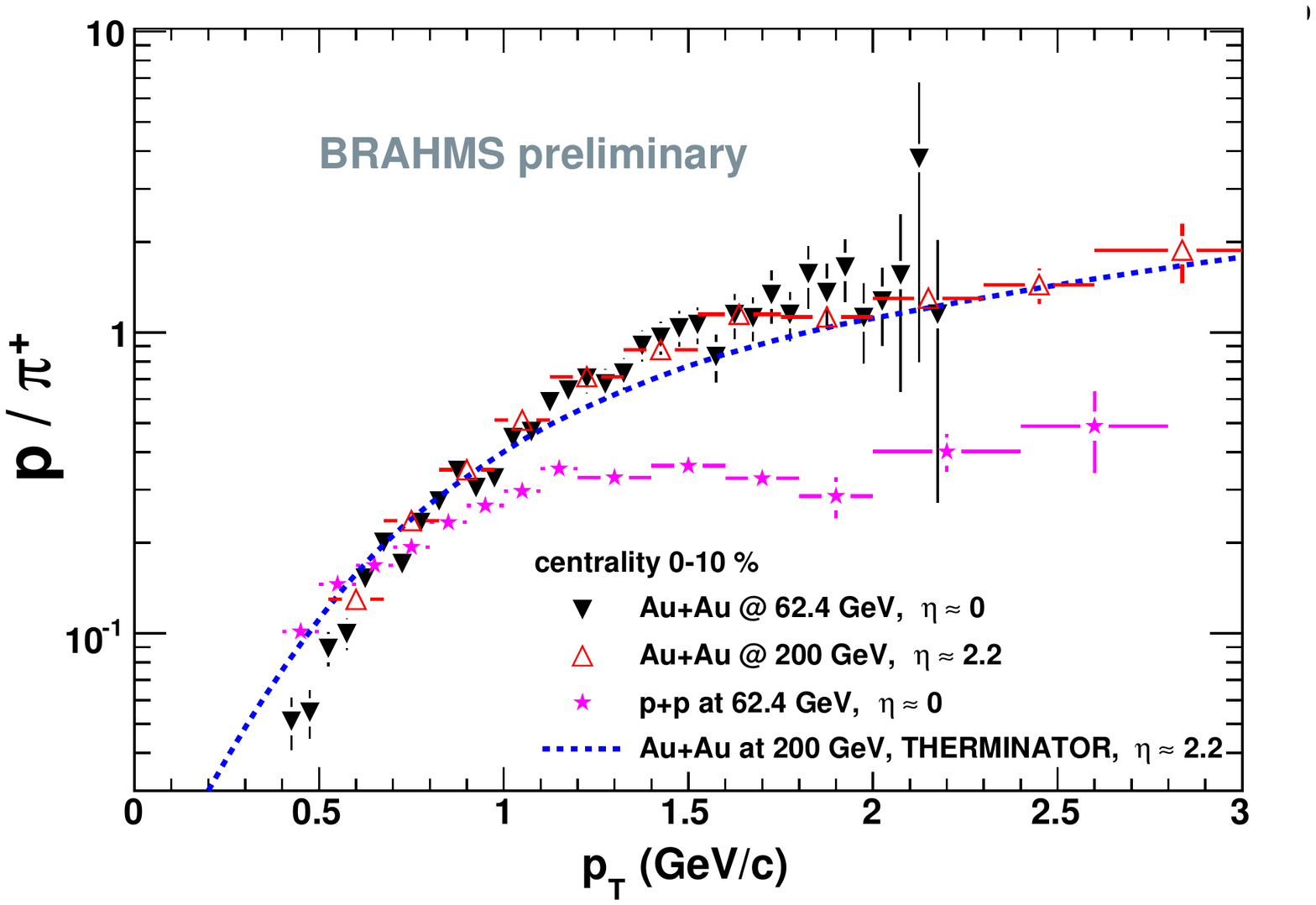}
\end{center}
\vspace{-0.8cm}
\caption{(color online) p/$\pi$ ratios vs.\pT~for central Au+Au collisions at $\eta=0$ and $\eta=2.2$ at approximately same $\bar{p}/p$ ratio. }
\label{fig:ppi}
\end{figure}
Figure \ref{fig:ppi} shows a comparison between the \ptopi measured in Au+Au collisions 
at \sNN=62.4 GeV and $\eta$ = 0.0 shown with closed black 
triangles and the same ratio measured in Au+Au reactions at \sNN = 200 GeV and  $\eta$ = 2.2 shown with the open red
triangles.
The pseudo-rapidity intervals at the two energies selected for this comparison correspond
to similar ${\rm \bar{p}/p} = 0.45$, which in turn, can be  connected to a common value of the baryo-chemical 
potential $\mu_{B}$ of the observed bulk media, equal to $\approx 62$ MeV  for these  two energies ~\cite{ratio200,ratio62}.
The lower values depicted by the grey stars show the \ptopip
ratio measured in the p+p system at \roots{62.4}. 
The similarity of proton-to-pion ratios for these selected
heavy ions collisions suggests that the baryon and meson production at the $p_{T}$ interval studied (up to 2 GeV/c) is dominated by medium effects and is determined by
the bulk medium properties. 
These strong medium effects are also suggested by
the observed enhancement of the p/${\rm \pi}$ as function of \pt~in the nucleus-nucleus
systems compared to the ratio extracted from nucleon-nucleon interactions.
In addition, Fig.~\ref{fig:ppi} shows that the THERMINATOR model ~\cite{THERM} calculations (dashed curve) describes central Au+Au data reasonably well.
This model is a 1+1 D hydromodel that incorporates rapidity dependence of statistical particle production (including excited resonances) imposed on the hydro-dynamical flow. Additional details and discussion can be found in another contribution to this conference~\cite{QM09PS}.

\section{Nuclear Suppression at High Rapidity in  d+Au Collisions}

The RHIC run-3 with d+Au was designed primarily to determine if the large suppression seen at intermediate to high \pT~ in meson spectra relative to scaled yield of p+p at mid-rapidity is an initial or a final state effect.
\begin{figure}[ht]
\begin{center}
\includegraphics[width=2.4in,angle=90]{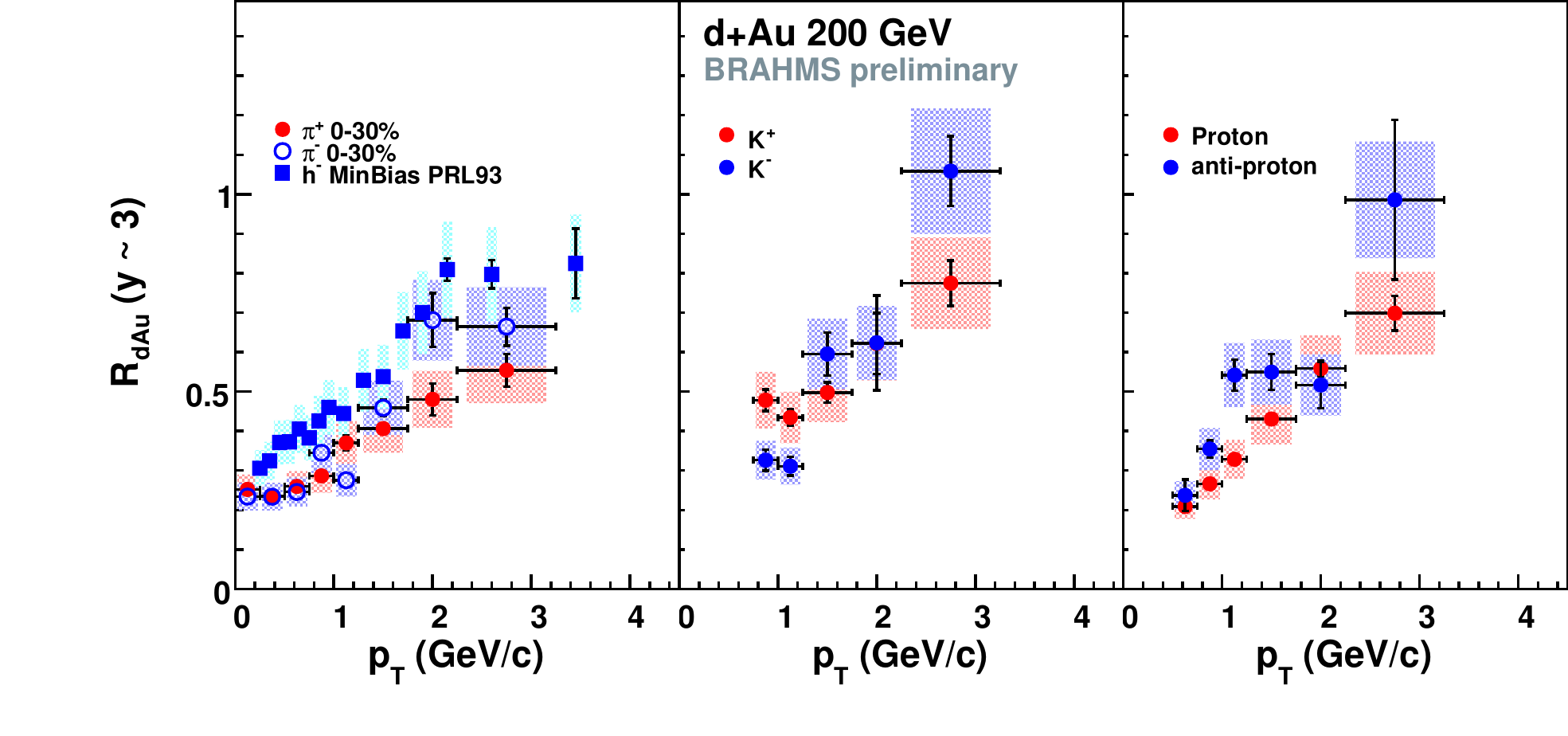}
\end{center}
\vspace{-0.5cm}
\caption{(color online) Nuclear modification factor identified $\pi$, $K$ and p in d+Au collisions at 200 GeV.}
\label{fig:RdA}
\end{figure}
The data
determined that final-state effects are in play, and are caused by partonic interactions of the hot and dense media formed in A+A collisions at RHIC energies.
It was also conjectured that measurements at forward rapidities could provide evidence for a new kinematic domain, where the high gluon density in nuclei would saturate and cause a reduction in the yield of produced particles. 
This state of matter was named the Color Glass Condensate (CGC).
The BRAHMS data provide an excellent testing ground for this idea by providing the nuclear modification factors $R_{\rm dAu}$ vs. rapidity for charged hadrons~\cite{BRAHMS:RdA}. 
The dependence of the data as a function of rapidity and centrality closely followed the  predicted signature for the CGC, albeit other explanations cannot be ruled out.
Here we present preliminary data for $R_{\rm dAu}$ for identified $\pi$, $K$, and protons at rapidity 3.2.
Figure~\ref{fig:RdA} shows that for the centrality bin $0-30\%$ all the identified hadron species exhibit a similar suppression pattern vs. \pt. 
The pions show a difference between \piplus~ and \piminus~ that can be attributed to the isospin dependence of the reference p+p pion yields at forward rapidities. 
It also demonstrates that the $h^-$ suppression  essentially equals that of \piminus~ and to those of other identified hadrons.

The observation of this suppression at high rapidity has been a subject of many theoretical investigations since the data were published, and I will make no attempt to properly reference all of the published papers.
Important modifications to the spectra can be expected from nuclear shadowing at small-$x$, kinematic suppression at large $x_F$, as well as other possibilities.
At this point, no definitive statement on the importance of Gluon Saturation at RHIC energies can be made and further result of correlation measurements are eagerly awaited.

\section{Summary}

In summary we highlight some specific lessons that have been learned from the BRAHMS data. 

\begin{itemize}

\item The net-proton distributions in peripheral Au+Au collisions have a similar shape as p+p.
A clear change in net-proton rapidity shape takes place at $\sim 60\%$ centrality from where on  distributions at more central collisions exhibit more stopping, with the bulk of the Au nucleus participating in the interaction.
The mean rapidity loss $\delta y$ is about 2 and almost constant above SPS energies~\cite{BRAHMS:WhitePaper,BRAHMS:stop62}.

\item The near Gaussian shape of produced mesons was a surprise, with the distributions bearing close resemblance to the prediction of the Landau hydrodynamical model, though it does not prove its validity~\cite{BRAHMS:WhitePaper,BRAHMS:auaumeson}.
\item The baryon chemical potential $\mu_B$ is a driving physics variable for many inclusive and  bulk observables such as particle ratios vs. y and  \pT~\cite{ratio200,ratio62}. In this proceeding we have shown that even the \pT-dependence of particle ratio's like $p/\pi$~ vs. \pt~ is governed in large part by this variable.
\item The differential elliptic flow decreases at forward rapidity, with the decrease for central events consistent with the expectations of 3D Hydro+Cascade calculations. Mid-central collisions at forward rapidity $v_2(p_T)$ show somewhat larger decrease towards forward rapidities. 
At forward rapidities a decrease in the mean \pt-values compared to mid-rapidity is observed, in particular for protons.
This is consistent with a reduction in radial flow at forward rapidity.


\item The d+Au high-\pt~ suppression observed at high rapidity is relevant for importance of the Color Glass Condensate and low-$x$ physics at RHIC energies and has inspired other new instrumentation and measurements in the forward region at RHIC~\cite{BRAHMS:RdA}.

\end{itemize}

Overall the BRAHMS experiment has provided unique physics results, particular in the forward region. Some of these questions are  ''how does matter behave at very high temperature and/or density'', ''what is the nature of gluonic matter'', and '' what is the  structure of the proton''.

\end{document}